\documentclass{iopart}
\usepackage[pdftex]{graphicx}
\begin{document}

\title{Electron Drift Directions in Strong-Field Double Ionization of Atoms}

\author{S L Haan$^1$, Z S Smith$^1$ K N Shomsky$^1$ P W Plantinga$^1$}

\address{$^1$Department of Physics and Astronomy, Calvin College, \\
Grand Rapids MI 49546, USA}
\ead{haan@calvin.edu}

\date{\today}

\begin{abstract}
Longitudinal momentum spectra and electron drift directions are considered for several laser wavelengths in Non-Sequential Double Ionization of helium using three dimensional classical ensembles.  In this model, the familiar doublet for wavelength 800 nm and intensities of order 5x10$^{14}$ W cm$^{-2}$,  becomes a triplet for wavelength 1314 nm, then a doublet with plateau for 2017 nm.  The results are explained based on whether the post-ionization impulse from the laser results in backward drift for one or both electrons.

Submitted to J Phys B on 30Dec08.
\end{abstract}

\maketitle

\section{Introduction}

One signature of Non-Sequential Double Ionization (NSDI) of atoms by intense laser fields \cite{reviews} is a double hump in the net or sum longitudinal{\footnote{The longitudinal direction is defined to be along the laser polarization axis}} momentum spectrum of the DI pairs \cite{doublet}.  The doublet is generally accepted to occur because the recollision process \cite{recollision} that drives the DI combined with the post-ionization impulse from the oscillating laser field \cite{Volkovkick} very often lead to electron pairs that drift out on the same side of the nucleus\cite{ChenandNam},\cite{Z-NZ}.  
By contrast, sequential ionization, in which the electrons ionize independently, gives uncorrelated electrons and  a single narrow peak in the net momentum.  Recollision Excitation with Subsequent Ionization(RESI) \cite{RESI} can also lead to uncorrelated electrons and a central peak\cite{Z-NZ},\cite{randomkick}, especially if there is long time delay between the recollision and the final ionization. 

If it weren't for the effects of the laser field or nucleus, then recollision impact ionization would lead to two electrons traveling forward, i.e., with longitudinal momentum parallel that of the recolliding electron just before the collision.  However, depending on laser phase at recollision and the electrons' speeds, the laser may push them back so that they drift into the backward direction\cite{ChenandNam}.  Reference \cite{PRL2006} 
showed that for classical ensemble models and laser parameters in the vicinity of wavelength $\lambda$=780 nm and intensities on the order of 5x10$^{14}$W cm$^{-2}$, the two electrons most often drift into the backward direction.  Reference \cite{PRL2006} also showed that recollision impact ionization occurred in only a minority of the trajectories, even though the returning electron could have enough energy for impact ionization.  It was more common for the recollisions to result in one free electron and one excited--but nonetheless bound--electron.  The free electron would be pushed into the backward direction by the laser field.  The bound electron would be pulled back by the nucleus and most often escape over the barrier and into the backward direction at the first laser maximum after recollision, often about one quarter cycle after recollision.  Reference \cite{ZachPRA} dubbed the latter process the ``boomerang."  If the electron that is bound just after the collision escapes before the field maximum, then to first approximation, it can be expected to drift into its direction of escape, which is backward relative to the recollision. Thus both recollision impact ionization and recollision excitation with the boomerang can lead to same-hemisphere electrons\cite{PRL2008}. However, excited electrons that escape over the barrier too late in the laser cycle (to first approximation, after the field maximum) drift opposite from their initial escape, thus into the forward direction relative to the recollision and opposite from the other electron. 

It's also possible for a free electron to scatter off the nucleus just before or just after recollision.  Such backscattering has been shown to be the source of electrons with energy above 2U$_p$ in NSDI \cite{ZachPRA},\cite{PRL2008},\cite{hotelectrons},\cite{parker390},\cite{outsidebox-expt},\cite{outsidebox-thy}.  Here U$_p$ denotes the ponderomotive energy, $E_0^2/(4\omega ^2)$, where $E_0$ is the laser field amplitude and $\omega$ the frequency.  (We use atomic units unless specifically indicated otherwise.)  2U$_p$ is the maximum drift energy for an electron that starts from rest in an oscillating electric field. At shorter wavelengths, such as 390 nm, high-energy electrons can also be produced through the boomerang\cite{ZachPRA}.

In this paper we consider the production of backward and forward drifting electrons, and situations other than RESI under which recollision can lead to oppositely directed electrons in NSDI.  We employ 3d fully classical ensembles as in Refs.~\cite{PRL2006},\cite{ZachPRA},\cite{PRL2008}, and \cite{JPB2008}.  We systematically adjust the laser wavelength as in the experimental work of Alnaser {\it et al}\cite{wavelengths} for argon and neon, thus changing U$_p$ and the recollision energy without having to change the laser intensity.

Ensemble size is typically one million.  To stabilize our starting state we soften the Coulomb potential, replacing $-2/r$ with $-2/\sqrt{r^2+a^2}$ where $a$=0.825.  Each two-electron trajectory begins with energy equal to that of the helium ground state and with each electron having zero angular momentum.  The starting ensemble is spherically symmetric. Each atom is allowed to propagate for a time equivalent to one laser cycle both before and after the laser pulse.  Each two-electron trajectory is found numerically by numerical integration of Newton's second law.  The laser is treated as an oscillating laser field that is uniform in space.  The first ionization occurs as a result of barrier suppression and e-e repulsion--there is no tunneling in this model.  After one electron achieves $|z|>$10 (where the z axis is the laser polarization axis), we change the shielding parameter to $a$=0.4, as described in \cite{ZachPRA} and  \cite{PRL2008}.  To conserve energy when we change $a$, we give each electron an appropriate kinetic energy boost in its radial motion.  This change of $a$ is necessary in order to have large-angle electron scattering off the nucleus at recollision.  Electron-electron shielding is kept constant at 0.05.  We use trapezoidal pulses, which have the characteristic of giving no net kick to the electron during turn on or turn off of the laser.  

In our analysis, we examined each doubly ionizing trajectory every 0.01 cycle and identified the final ionization times for each electron.  We define an electron to be ionized if it achieves, and then maintains for at least 0.2 cycles, any of: $E>0$, where E is its energy, inclusive of electron-nucleus and e-e interactions but not the laser interaction; $|z|>10$; or $zF_z>0$  with $z^2>5$ where $F_z$ is the longitudinal component of the net force.  The final test basically checks whether the net force on the electron is toward or away from the nucleus, an indicator of whether particle is inside or outside the nuclear well.  We then scan the time interval from when one electron first achieves $|z|>10$ until final ionization of both electrons, and we call the time of closest approach the recollision time.

\begin{figure}
\centerline{
{\includegraphics[scale=0.4]{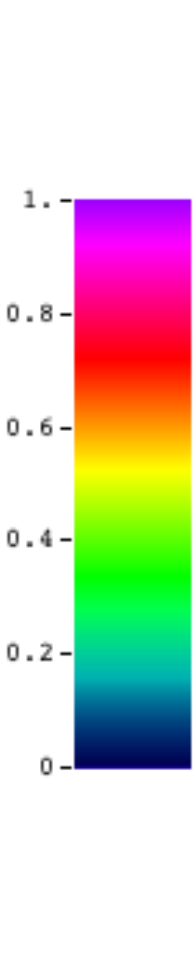}}
{\includegraphics[scale=0.4]{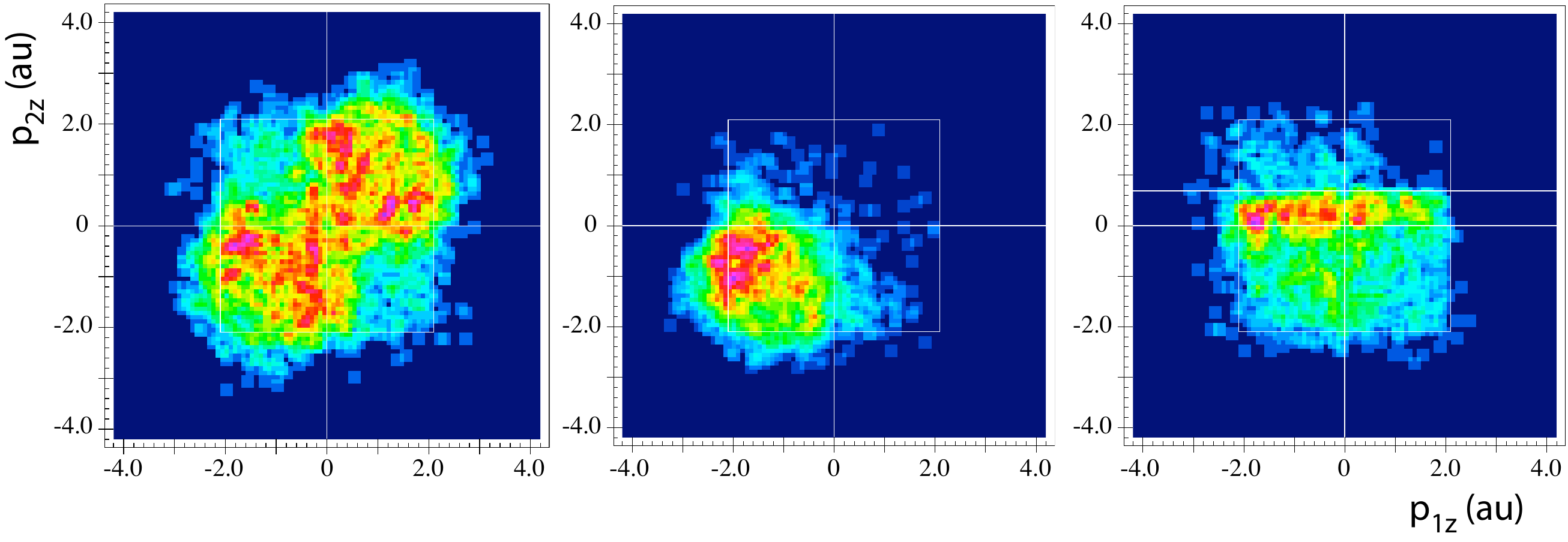}}
}
\caption{Final longitudinal momentum for one DI electron vs the other for $\lambda$=800 nm, I=0.5 PW cm$^{-2}$ for a 10 cycle (2+6+2) pulse, with softening parameter $a$=0.4.  Left:  No distinction between electrons.  Center and right:  recolliding electron vs struck electron.
In the center plot we show trajectories with time delay up to 0.25 cycles, on the right those with time delay above 0.25 cycles.  The horizontal line in the right-hand plot is explained later.  Ensemble size is 800K, giving 4813 double ionizations.}
\label{800I5momenta}
\end{figure}

\section{Wavelength $\lambda$=800 nm}

We consider first wavelength $\lambda$=800 nm and laser intensity 5x10$^{14}$W cm$^{-2}$ ($U_p$=1.10).  
In Fig.~\ref{800I5momenta} we show final longitudinal momentum of one DI electron vs the other.  Pulse length is 10 cycles (2 cycle turn-on +6 cycles full strength +2 cycle turn-off), and ensemble size is 800,000.  The boxes show momentum $2\sqrt{U_p}$.  On the left, we make no distinction between the electrons.  Population extends beyond the box, but the sum momentum $|p_{1z}+p_{2z}|$ has maximum value close to $4\sqrt{U_p}$.  In the center and right-hand plots we define the direction of recollision (the longitudinal direction of motion of the returning electron just before recollision) as positive, and we plot the final longitudinal momentum of the recolliding electron vs that of the struck electron.  The center plot includes trajectories with time delay, from recollision to final ionization, of up to 0.25 cycles.  Most of the population lies in the third quadrant, indicating having both electrons drift into the backward direction relative to the recollision.  The third plot considers trajectories that have time delay more than 0.25 cycles, and shows considerable population in the third and fourth quadrants.  The most conspicuous feature of the plot may be the band just above the $p_{2z}$ axis.  Here the recolliding electron drifts into the forward direction relative to the recollision, up to a certain cutoff momentum.  Below, we interpret these in terms of recollision-excitation trajectories in which the free electron has sufficient energy after the collision to overcome the push back from the laser and drift into the forward direction. 
   
\begin{figure}
\centerline{
{\includegraphics[scale=0.55]{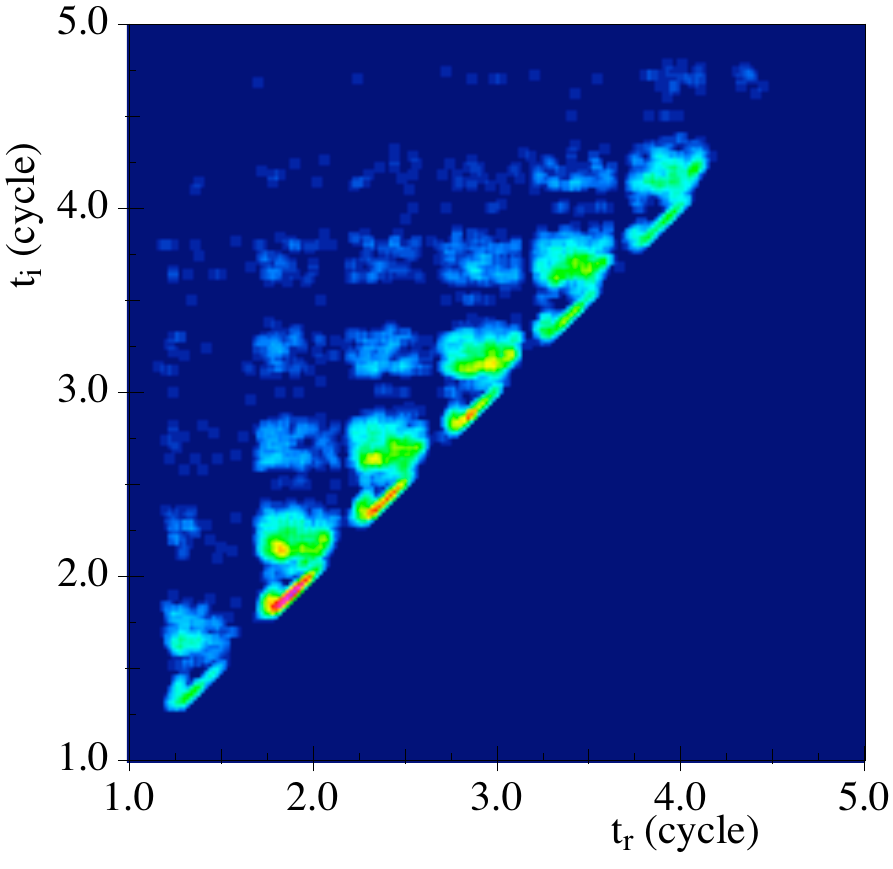}}
{\includegraphics[scale=0.54]{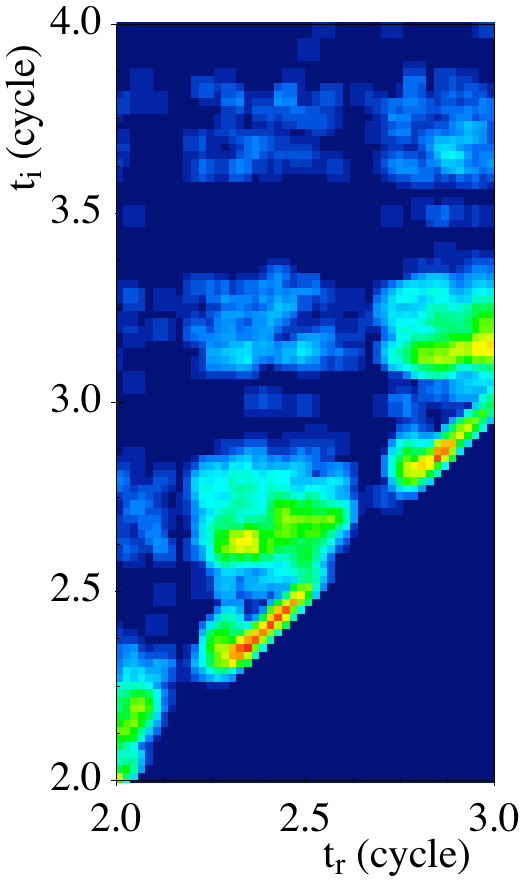}}
{\includegraphics[scale=0.55]{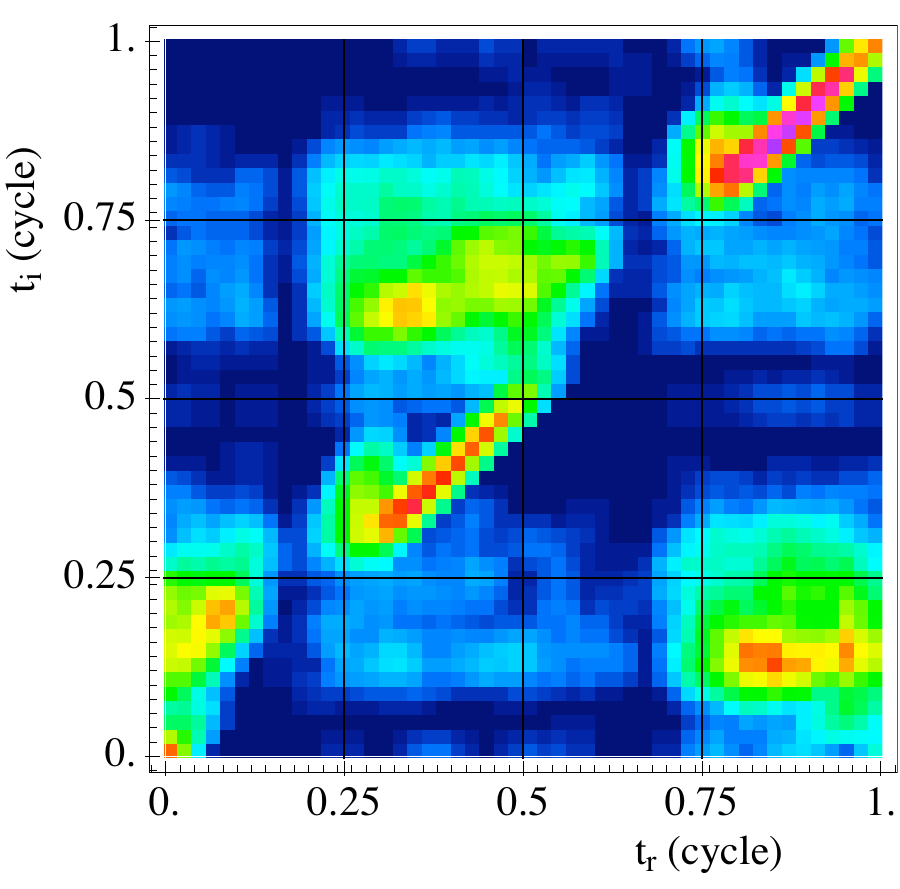}}
}
\caption{
Time of final ionization vs time of recollision for all DI trajectories in a 5 cycle pulse for $\lambda=800 nm$, I=5 x $10^{14}W cm^{-2}$, and a=0.4.  In the center we zoom in one cycle.  On the right we allow for wraparound to collapse all data to 1 cycle.}
\label{phaseplotI800}
\end{figure}

In the leftmost plot of Fig.~\ref{phaseplotI800} we plot final ionization time vs recollision time for a 5-cycle (1+3+1) pulse.   Impact DI is indicated by population along the diagonal.  Because of the one-cycle laser turn on, recollisions do not begin until about 1.25 c and don't reach maximum energy until the interval from 1.75 c to 2 c.  Other DI populations can be associated with RESI.  Considerable ionization is evident at the first laser maximum after recollision,  with decreasing amounts at subsequent maxima.  

Because of the difficulty in seeing details in the leftmost plot, we zoom in on recollision time from 2 to 3 c in the center plot of Fig.~\ref{phaseplotI800}.    In the right plot we include all DI pairs and allow for wraparound, plotting laser phase at final ionization vs laser phase at recollision (denoted by $t_i$ and $t_r$ respectively, and measured in laser cycles). 
The plots show that recollision impact ionization occurs in time intervals from just after peak field until the field zero.  There is a clear ``shadow" of slightly delayed ionization while the field is strong.  Also clearly evident is DI from slowdown collisions, in which recollision occurs after the field zero so that the returning electron is traveling against the laser force.  Such collisions were very important in 1d  \cite{panfili} where there was, in effect, only one impact parameter for the recollisions and it was important to match the motions of the returning and bound electrons so as to maximize energy transfer.  In 3d, electrons return with a variety of impact parameters, and a wider range of recollision times is effective. 
 
\begin{figure}
\centerline{
{\includegraphics[scale=0.6]{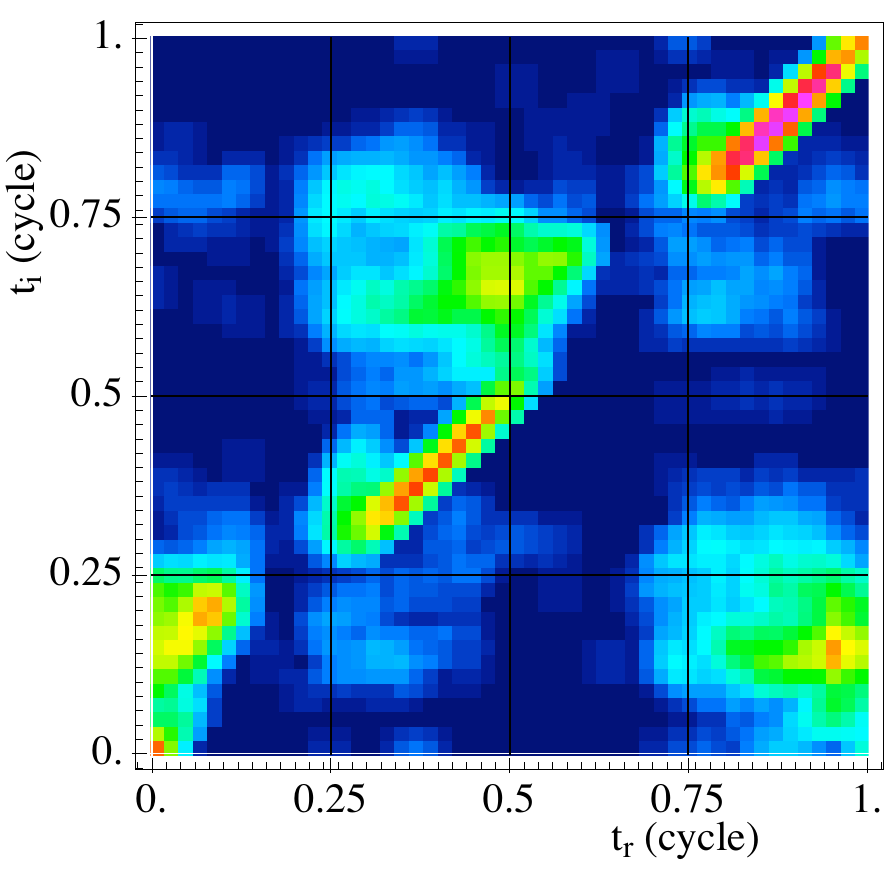}}
{\includegraphics[scale=0.6]{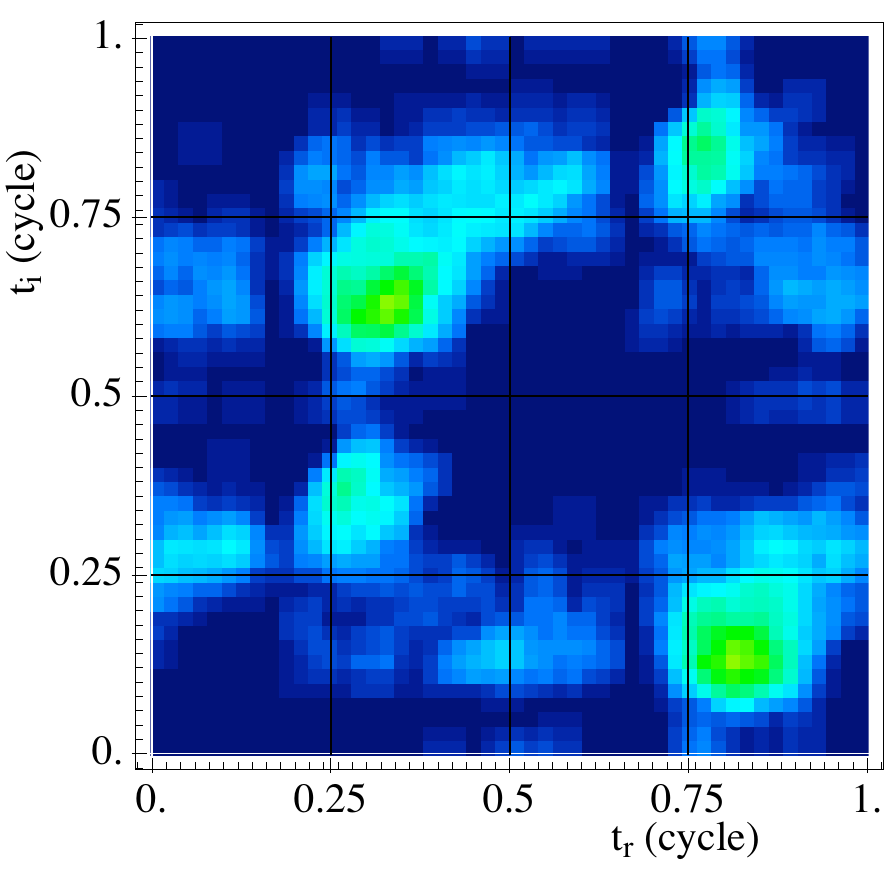}}
}
\caption{Similar to rightmost plot of Fig.~\ref{phaseplotI800}, but with trajectories separated based on whether the DI electron pairs drift out in the same (left) or opposite (right) longitudinal directions.  Color scale is the same for both plots, but independent from Fig.~\ref{phaseplotI800}.}
\label{800phaseplotbyhemi}
\end{figure}

In Fig. \ref{800phaseplotbyhemi} we divide the population of Fig.~\ref{phaseplotI800} into two parts, based on whether the electrons drift out with the same or opposite signs for final $p_z$.   We shall occasionally refer to these two cases as emerging in the same or opposite hemispheres.  
The plot on the left, for same-hemisphere electrons, is a reminder that recollision excitation with ionization before the next field maximum can produce two electrons that drift out together as a correlated pair\cite{PRL2008}, often with one electron pushed back by the laser and the other boomeranging.  
In the right-hand plot, population for time periods in which the laser field is waning ($t_i$ from .25c to .50c and again .75 c to 1.00c) is a reminder that anticorrelated electrons can be produced if final escape occurs after the field maximum\cite{PRL2006}.  Clusters in the right-hand plot near $(t_r,t_i)=(0.32,0.65)$ and $(0.82,0.15)$ indicate oppositely directed electrons even though final emission occurs before the field maximum. These have time delay greater than 0.25 cycle, and thus were included in the right plot of Fig.~\ref{800I5momenta}.  We interpret these in terms of the recollision-excitation trajectories in which the free electron has sufficient energy after the collision to drift into the forward direction, as we discuss below. 

\begin{figure}
\centerline{
{\includegraphics[scale=.25]{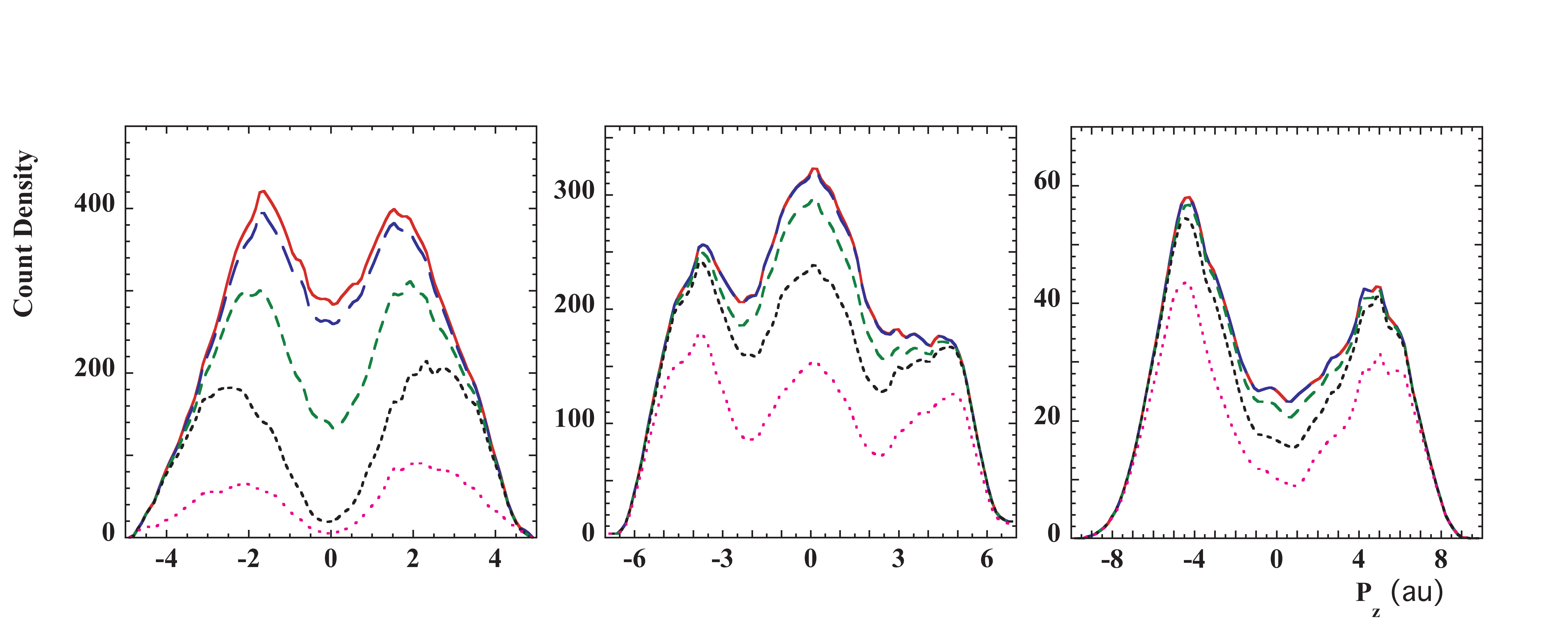}}
}
\caption{Spectrum of net longitudinal momentum for I=5x10$^{14}$W cm$^{-2}$ and for $\lambda$=800 nm (left), 1314 nm (center), and 2017 nm (right) for 5-cycle pulses.  Maximum time delays from bottom to top are 0.06 c, 0.26 c, 0.50 c , 2.00 c.  Top curves show full spectra through the end of the pulse.}
\label{netmomenta}
\end{figure}

We consider next the net or sum longitudinal momentum, $p_{1z}+p_{2z}.$  On the left in Fig.~\ref{netmomenta} we show the net longitudinal momentum spectrum for increasing time delay between recollision and final ionization.
As shown in  Fig. 5 of Ref.~\cite{PRL2008} for $\lambda$ = 780 nm and intensity 4x10$^{14}$W cm$^{-2}$, a doublet forms for time delays of a portion of a laser cycle, then fills in.  The center and right plots apply for wavelengths 1314 and 2017 nm, and will be discussed below.

\section{Wavelength $\lambda$=1314 nm}

Changing the laser wavelength but maintaining the same laser intensity changes the ponderomotive energy $U_p$  
and thus the energy available at recollision.  Elsewhere \cite{JPB2008}, we have considered the smaller wavelength $\lambda$=483 nm.  For that case, the net momentum spectrum was a singlet and the most common route to DI was recollision with a short-lived doubly excited state.  In the present work we consider longer wavelengths, so that the recollision energy is increased.  Pulse length remains five cycles. 

\begin{figure}
\centerline{
{\includegraphics[scale=0.72]{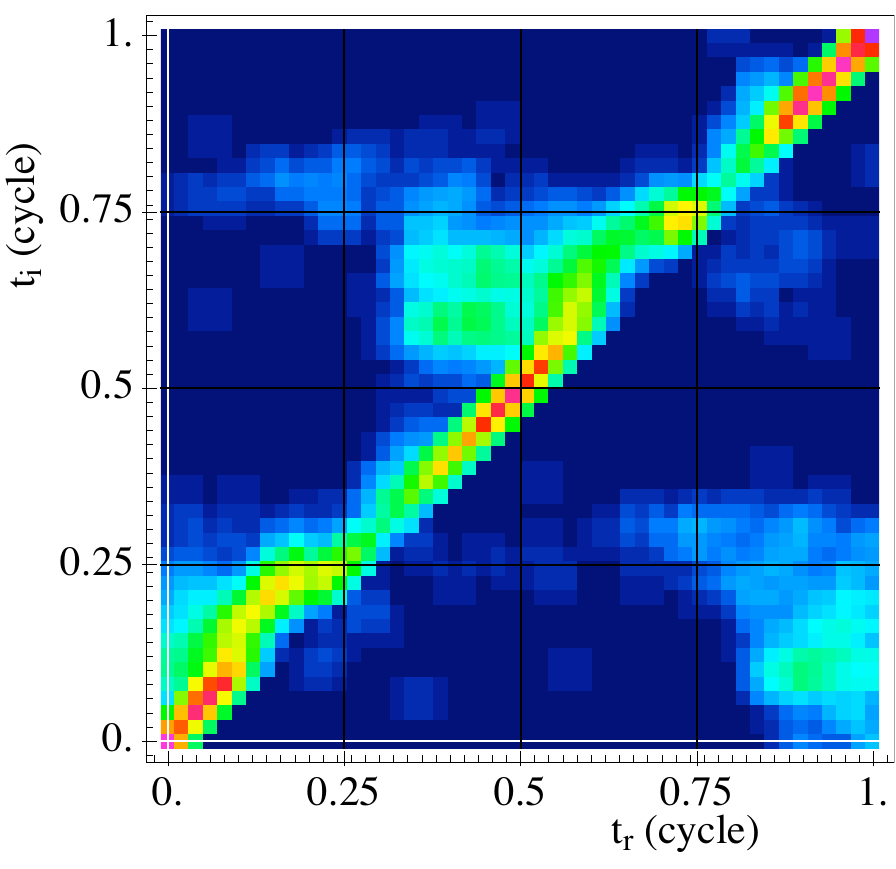}}
{\includegraphics[scale=0.72]{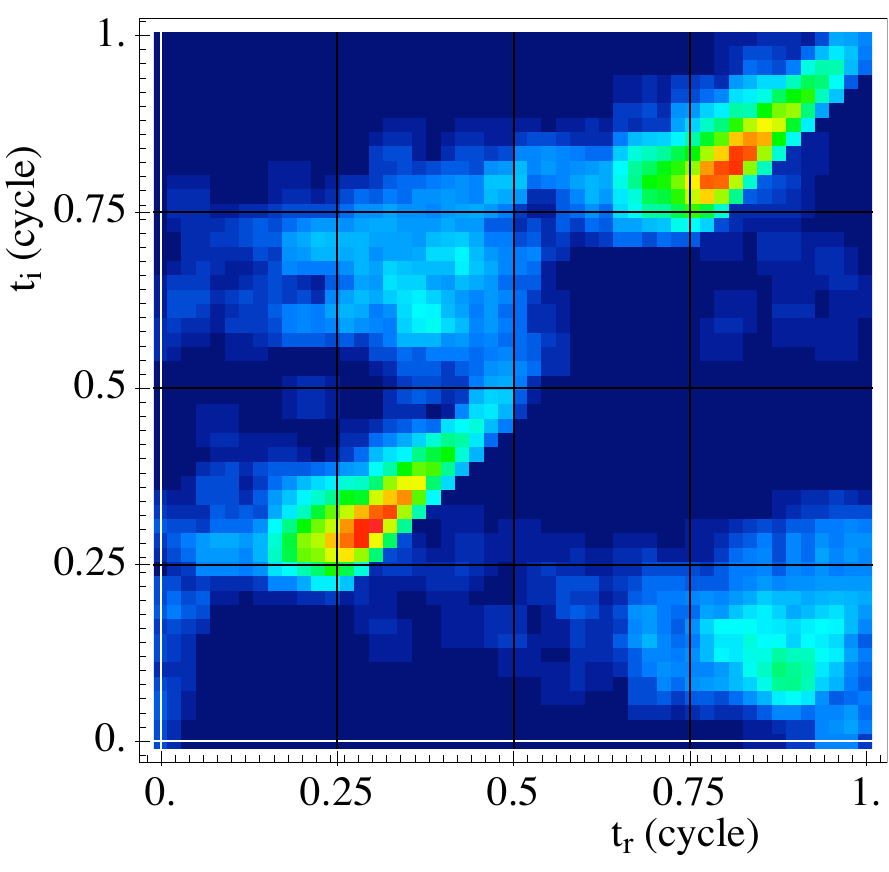}}
}
\caption{Phase plots for $\lambda$=1314 nm, I=5 x 10$^{14}$ W cm$^{-2}$.  Same-side trajectories are included on left, opposite-side on right.  The right-hand plot indicates that recollision ionization shortly after the field maxima can lead to opposite-hemisphere electrons.}
\label{1314phase2hemi}
\end{figure}

Phase plots for same-hemisphere and opposite-hemisphere electrons for $\lambda$=1314 nm ($U_p$=2.97) are shown in Fig.~\ref{1314phase2hemi}.  For these laser parameters, there is considerable population along the diagonal, indicating increased importance of impact ionization at recollision. We've determined the median time delay from recollision to final ionization to be 0.06 cycles, so about half the DI can be attributed to recollision impact ionization.    
The right-hand plot reveals that recollisions that occur shortly after the field maxima (0.25 and 0.75 c) can lead to significant numbers of opposite-hemisphere electrons.  Because of these oppositely directed electrons the net momentum spectrum is a triplet, as shown in the center plot of Fig.~\ref{netmomenta}.  The existence of the triplet for even short time delay indicates that the triplet is not the result of RESI.  Nonetheless, it does arise from opposite-hemisphere electrons.  The left-right asymmetry in the net momentum spectrum arises because collisions that favor the left peak occur first in the pulse.

\begin{figure}
\centerline
{\includegraphics[scale=0.50]{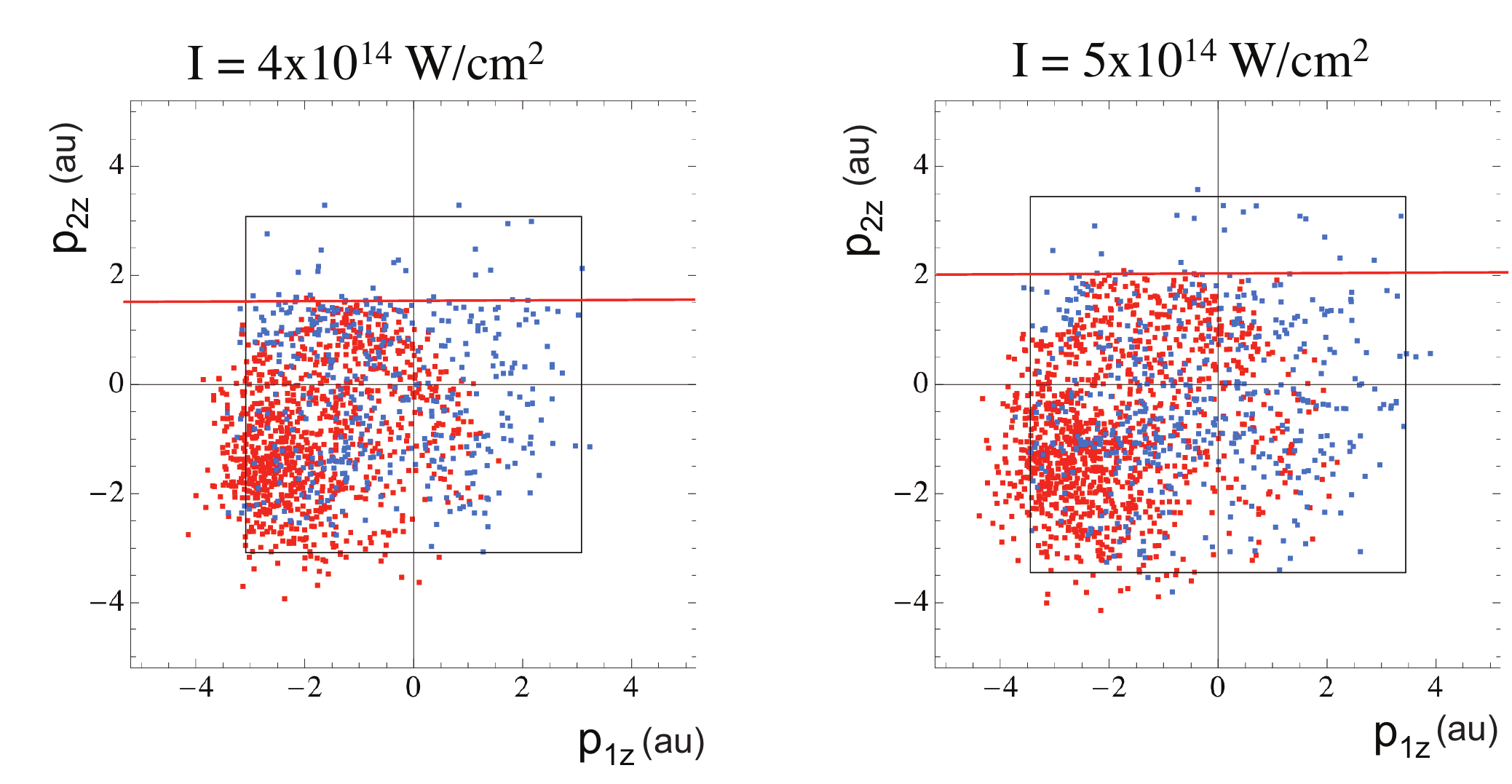}}
\caption{Scatterplot for $\lambda$=1314 nm and two intensities 4 and 5 x 10$^{14}$ W cm$^{-2}$.  Trajectories are distinguished red vs blue based on whether time delay is less than or greater than 0.25 cycle, respectively.  The horizontal lines show sharp cutoffs for forward propagation.  Boxes indicate momentum 2$\sqrt{U_p}$. 
 }
\label{1314scatter}
\end{figure}

In Fig.~\ref{1314scatter}, we show scatterplots of longitudinal momentum $p_z$ for intensities 4 and 5 x 10$^{14}$ W cm$^{-2}$ and for the recolliding electron (vertical axis) vs the struck electron (horizontal axis), with, for each DI pair, the forward direction defined as positive.  Red dots indicate trajectories with time delay (from recollision to final ionization) up to 0.25 cycle, and blue dots the trajectories with longer time delay.
The top-left quadrant includes trajectories in which the recolliding electron continues in the forward direction but the struck electron drifts into the backward direction, either through boomeranging or pushback by the laser.  As in Fig.~\ref{800I5momenta}, 
there is also a clear cutoff for the final momentum of the forward-drifting electron, marked by the horizontal lines.  As for $\lambda$=800 nm, we interpret the forwardly directed electrons in terms of recollision-excitation trajectories in which the recolliding electron has sufficient energy after the collision to drift into the forward direction.  Because it must deliver enough energy for the other electron to escape, there is a sharp energy cutoff.  The blue dots in Fig.~\ref{1314scatter} can be associated with RESI; of course, in our model all ionization is over the barrier.  The second electron may emerge in either direction, with momentum up to 2$\sqrt{U_p}$, indicated by the boxes.

\section{The Forward Drift}

We consider here conditions under which recollision can immediately result in a free electron that drifts into the forward direction.  We employ the standard three step model of (1) initial ionization (time $t_0$), (2) acceleration by the laser field (with other forces ignored), and (3) recollision (time $t_r$).  The initial ionization time determines the energy $E_r$ that the recolliding electron has just before recollision.  Earlier initial ionizations correspond with later returns.

We treat the recollision as instantaneous, so the electron speed after recollision is 
\begin{equation}
v_{\phi}=[2(E_r-\Delta E)]^{1/2},
\end{equation}
where  $\Delta E$ denotes the energy that the electron gives up in the recollision and $\phi=\omega t_r$ indicates the laser phase the instant after recollision.\footnote{Due to electron exchange, $v_{\phi}$ may be the speed of either electron after the collision.}  If the motion is longitudinal (i.e., parallel to the laser polarization axis), the drift velocity can be approximated (neglecting forces other than from the laser) as 
\begin{equation}
v_d=v_{\phi}-2\sqrt{U_p}cos\phi.  
\end{equation}
Thus, for example, if recollision at the time of the laser zero ($\phi=2\pi$) were to result in an electron initially at rest ($v_\phi=0$), that electron would obtain drift velocity $-2\sqrt{U_p}$, with the minus sign indicating drift into the backward direction.
To examine the conditions under which $v_d$ can be positive, we consider the limiting case in which the recolliding electron only gives up enough energy that the other electron will be able to escape over the barrier at a subsequent field maximum.  For nuclear potential -2/r, the threshold energy for over-the-barrier escape is $-4\sqrt{\omega}U_p^{1/4}$.  Thus,  if the inner electron begins in the ionic ground state with energy $E_g$, the energy delivered must be at least
\begin{equation}
\Delta E_{min}=-4\sqrt{\omega}U_p^{1/4}-E_g.
\end{equation}
Using this minimum energy in Eq.~(1) for $v_{\phi}$ allows us to determine numerically the maximum final drift velocity as a function of initial ionization time $t_0$, laser frequency $\omega$, and ponderomotive energy $U_p$.  All three parameters are needed.  Also, any transverse velocity would imply decreased forward velocity $v_{\phi}$ immediately after the collision.  

\begin{figure}
\centerline{
{\includegraphics[scale=.7]{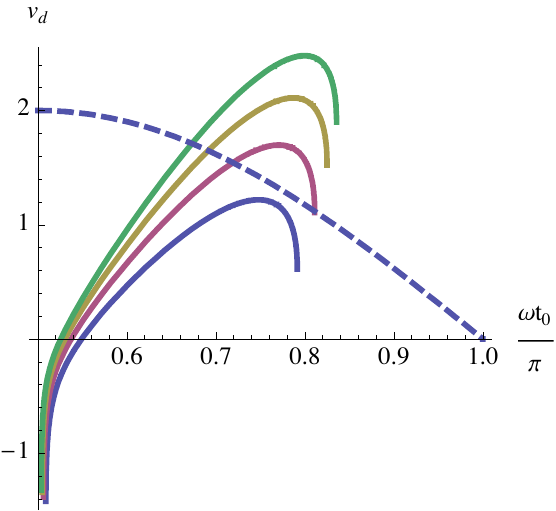}}
{\includegraphics[scale=.7]{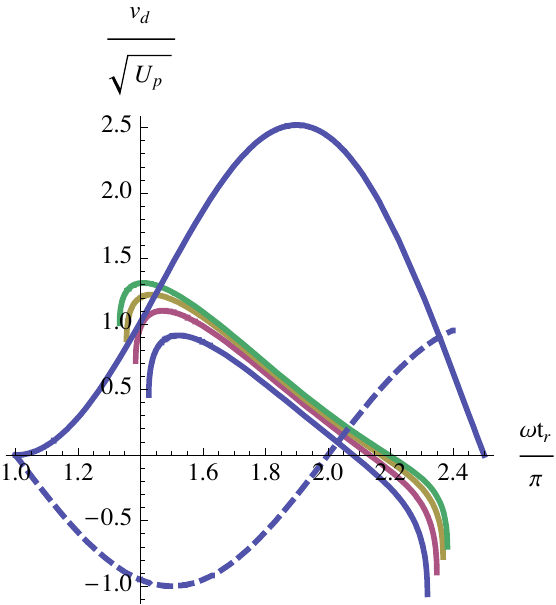}}
}
\caption{Calculated drift velocities for $\lambda$=1314 nm for intensities (from bottom to top) I=3,4,5,and 6 x 10$^{14}$ W cm$^{-2}$.  Left: velocities $v_d$ vs laser phase of first electron (initial) ionization.   Right:  $v_d/\sqrt{U_p}$  vs phase at recollision.  Dashed curves show laser field in arbitrary units, and the solid, topmost curve on the right shows electron velocity/$\sqrt{U_p}$ immediately before the collision.  Laser phases are expressed as multiples of $\pi$.  Values of $U_p$ are 1.78, 2.38, 2.97, and 3.56.}
\label{drifts1314}
\end{figure}

In Fig.~\ref{drifts1314}  we plot maximum forward drift velocities for $E_g=-2$, $\lambda$=1314 nm, and intensities 3, 4, 5, and 6 x 10$^{14}$ W cm$^{-2}$. On the left, we plot vs laser phase ($\omega t_0$) at initial ionization.  On the right, we divide the velocities by $\sqrt{U_p}$ and plot vs $\phi$, the laser phase at recollision.  The dashed curves indicate the laser field (drawn at arbitrary amplitude), and the topmost curve in the right-hand plot shows the returning electron's velocity (/$\sqrt{U_p}$) just before the collision.  The sharp cutoffs that the curves in the left plot show for larger $\omega t_0$ and, equivalently, that the curves in the right plot show for smaller $\phi$ are present because later initial emissions lead to earlier, less energetic returns, and the return energies become too small to excite the other electron to the threshold for subsequent escape.  However, earlier recollision times allow for increased forward acceleration by the laser field after the recollision.  Hence, the greatest forward drift velocities do not come from the most energetic recollisions, but from collisions closer in time to the laser maximum.  This result is consistent with what we saw in Fig.~\ref{1314phase2hemi}.   The maximum values for $v_d$ that we calculate for I=0.5 PW cm$^{-2}$  closely match the ensemble cutoff velocities shown in Fig.~\ref{1314scatter}.

\begin{figure}
\centerline{
{\includegraphics[scale=.7]{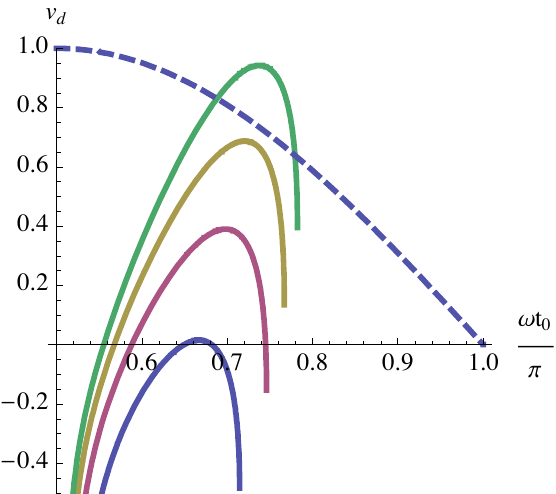}}
{\includegraphics[scale=.7]{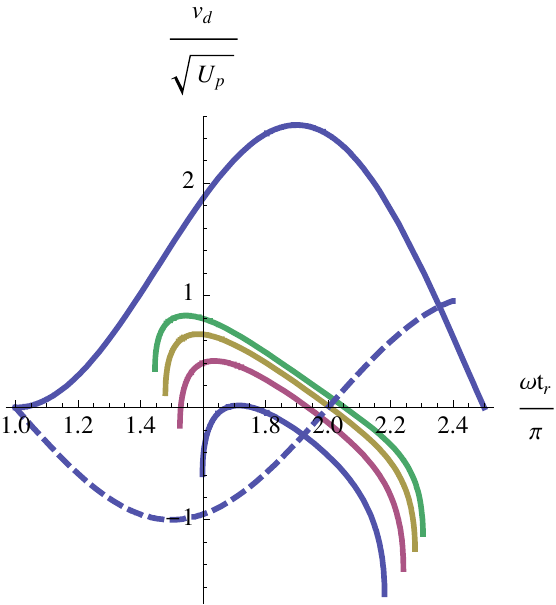}}
}
\caption{Electron drift velocities, repeating Fig.~\ref{drifts1314}, but for laser wavelength 800nm.  Values of $U_p$ are 0.66, 0.88, 1.10, and 1.32.}
\label{drifts}
\end{figure}

Figure \ref{drifts} repeats Fig.~\ref{drifts1314} in showing maximum drift velocities, but for $\lambda$=800 nm.  The threshold for being able to obtain a forward directed electron at recollision is reached at about I=3.0x10$^{14}$W cm$^{-2}$.

If the recollision leaves the other electron bound, that electron may boomerang, thus giving opposite-hemisphere electrons as in quadrant two of the rightmost plot of Fig.~\ref{800I5momenta}.  Delayed escape by the other electron could lead to its drifting in either direction, which explains the spillover into quadrant one of Fig.~\ref{800I5momenta}.

Rather than considering the threshold for recollision excitation, we can consider the threshold for direct ionization of the second electron.   Then the minimum energy that must be delivered is simply $\Delta E_{min} = -E_g$, which gives $v_{\phi}=[2(E_r+E_g)]^{1/2}$.  It is straightforward to show numerically that forward drift velocity can be obtained if  $U_p > 0.57|E_g|$, independent of $\omega$.  At  $U_p = 0.57|E_g|$ an initial ionization that occurs at $\omega t_0=0.67\pi$ or 120 degrees (0.33 cycle) leads to recollision at $\phi=\omega t_r=1.72\pi$ or 309 degrees (0.86 cycle) and an electron with forward velocity $v_\phi = 2\sqrt{U_p}cos(\phi)$ just after the collision, hence zero {\it drift} velocity.  The other electron would have zero velocity immediately after the collision and be pushed into the backward direction by the laser.  As $U_p$ increases above the threshold value, the range of original emissions that can lead to a forward drifting electron increases.  

As $U_p$ increases above the threshold, progressively more energy becomes available for the two electrons after recollision.  The next threshold occurs when both electrons can have sufficient forward velocity after collision to drift into the forward direction.  For this to occur both electrons need forward velocity $v_\phi$ exceeding $2\sqrt{U_p}cos(\phi)$ just after the collision.  It is straightforward to show that this can occur if $U_p>0.77|E_g|$  At the threshold for having two forward drifting electrons, the return time is 0.80 cycles or 1.60${\pi}$ (287 degrees) and original emission time is 0.36 c or .715$\pi$ (129 degrees).  Of course, near the threshold value the two electrons would need to share energy nearly equally for both to drift into the forward direction.  Unequal sharing would lead to electrons drifting out in opposite directions, just as we found in Fig.~\ref{1314phase2hemi} for $\lambda$=1314 nm.  Our experience has been that equal energy sharing is very unusual; hence we would expect that the system would need to be well above threshold before forward traveling pairs became common.  

\section{Laser Wavelength 2017 nm}

\begin{figure}
\centerline{
{\includegraphics[scale=0.72]{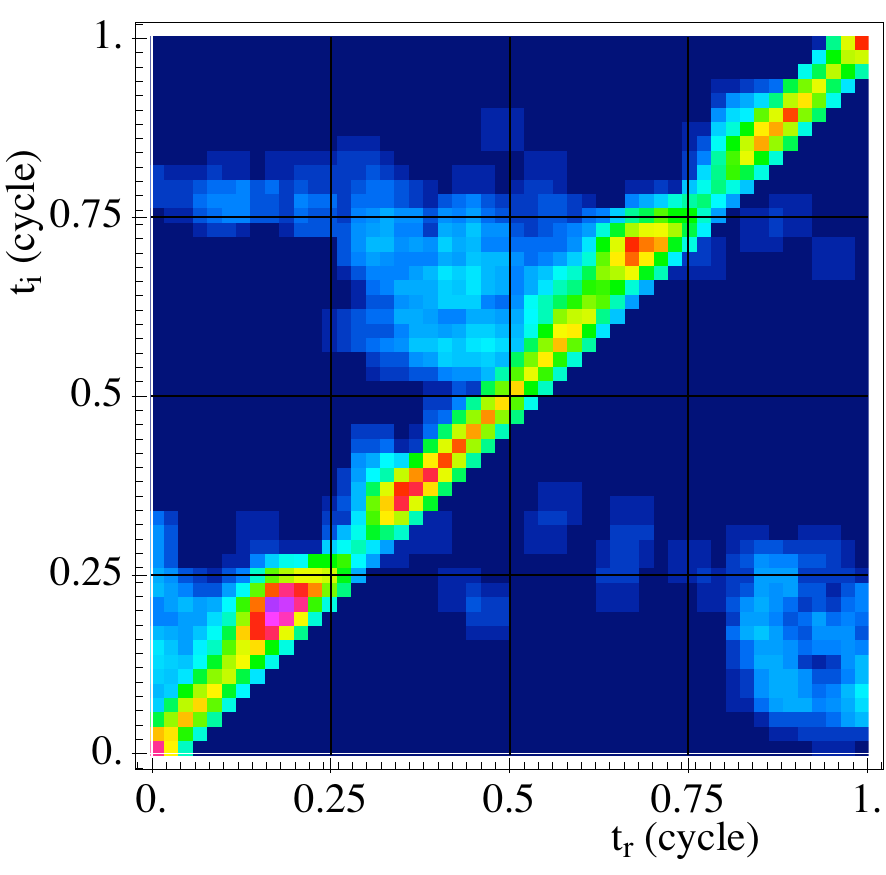}}
{\includegraphics[scale=0.72]{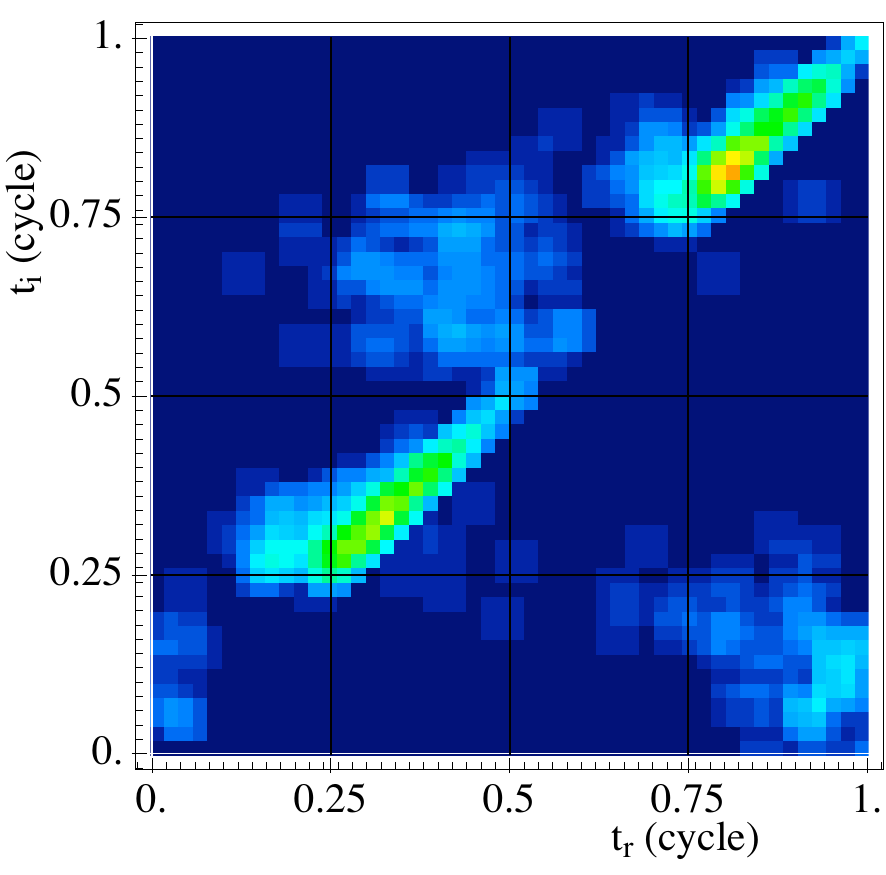}}
}
\caption{Phase plots for $\lambda$=2017 nm, I=5x10$^{14}$W cm$^{-2}$, for a 5-cycle (1+3+1) pulse.  Same-hemisphere are included on left and opposite-hemisphere trajectories on the right.}
\label{2017phaseplot}
\end{figure}

Figure \ref{2017phaseplot} shows final ionization phase vs recollision phase for wavelength 2017 nm and for the same intensity of 5x10$^{14}$W cm$^{-2}$.  It shows that recollision ionizations occur over a wide range of laser phases, including just before laser maxima (.25 and .75 c).  Having recollision ionization occur before the laser maximum assures forward drift.  Mathematically, we would have $cos\phi<0$ in Eq.~(2).  We can also obtain forward-forward pairs from other collision times, since we are above the threshold determined in the previous paragraph.  One can expect some variation in relative electron energies immediately after impact DI, with lower energy electrons pushed into the backward direction.  Thus, we obtain opposite-hemisphere electrons, as indicated in the right hand plot, as well as same hemisphere electrons.

As an aside, we note that we have smaller DI yield at this large wavelength.  The barrier is suppressed for much longer each half cycle.  Consequently, more first ionizations occur before the field maximum and a smaller fraction of electrons return for recollision.

The net longitudinal momentum spectrum is displayed in the rightmost plot of Fig.~\ref{netmomenta}.  It is again a doublet, but with a central plateau from the opposite-hemisphere electrons.  The doublet is so unequal because of preference for recollision early in the pulse.

\section{Conclusions}

We have examined how NSDI within classical models varies with laser wavelength.   Changing the wavelength but not intensity changes the ponderomotive energy $U_p$, and thus the energy at recollision.  When the wavelength is increased in our model the net or sum longitudinal momentum transitions from a singlet at wavelength 483 nm to a doublet at 800 nm, then a triplet at 1314 nm, then back to a doublet at 2017 nm.  The short wavelength case has been examined elsewhere  \cite{JPB2008}, where we discussed how recollision excitation could lead to a doubly excited state that would decay into oppositely traveling electrons.  At 780 or 800 nm, the most common scenario for NSDI is recollision excitation.  One electron remains free after the recollision and is swept into the backward direction by the laser.  The other electron is bound but often boomerangs \cite{ZachPRA} (is pulled back by the nucleus) and escapes into the backward direction at the first laser maximum after the recollision.  However, already at 3x10$^{14}$W cm$^{-2}$ we are well above the threshold for one electron to be able to retain enough energy at recollision excitation that it can overcome the backward push from the laser field and drift into the forward direction.  Such electrons are a small part of the total at 800 nm, but become much more important at 1314 nm.  Because the other electron is likely to be pushed back by the laser field, oppositely directed electrons can be obtained, giving rise to a central peak in the spectrum.  At this long wavelength, the outer peaks are well separated and the central peak distinct, so the spectrum becomes a triplet.  At still higher wavelengths we cross the threshold for having two forward directed electrons after recollision impact ionization.  That suppresses the center peak, so that the spectrum is again a doublet.

Recent experiments by Rudenko {\it et al} \cite{RudenkoPRA2008} have looked at variation of the momentum spectrum with laser intensity.  They have seen how the doublet collapses to a singlet as systems transition from NSDI to sequential ionization.  It may be that in the transitional region, the growing central spike is not just from sequential ionization, but from recollision generated forward-backward (or ``Z") combinations such as we have discussed here.  We expect that in order to see these combinations unambiguously, longer wavelengths would be needed.
 
In their experiments, Alnaser {\it et al} \cite{wavelengths} have seen the transition from doublet to singlet in the spectrum for Ne$^{2+}$ for increasing intensity at wavelength 1314 nm, but with no triplet.  We are continuing to investigate the species dependence of the effects we've been considering.  
Forward traveling electrons can be produced from recollisions that occur while the laser field is strong, but such recollisions occur in the three-step model only if first ionizations can occur fairly late in a laser pulse.  It may be that it is too difficult to ionize the first electron from neon (which has high binding energy) for the triplet to be seen there.

\ack{This material is based upon work supported by Calvin College and by the National Science Foundation under Grant No.~0653526 to Calvin College.  We acknowledge also our ongoing collaboration with J.H. Eberly's group at the University of Rochester. }


\section*{References}
\begin{thebibliography}{10}

\bibitem{reviews}For reviews, see 
D\"{o}rner R {\it et al}
2002 {\it Adv.  At. Mol. Opt. Phys.} {\bf 48} 1-34; 
Ullrich J {\it et al}
2003 {\it Rep. Prog. Phys.} {\bf 66} 1463-1545;
and
Becker A, D\"{o}rner R, and Moshammer R,
2005
{\it J.~Phys.~B: At.~Mol.~Opt.~Phys.}  
{\bf38} S753
-S772



\bibitem{doublet}See for example
Weber Th, {\it et al}
2000 {\it \PRL}{\bf 84} 443-446
and 
Moshammer R {\it et al}
2000 {\it \PRL}{\bf 84} 447-450

\bibitem{recollision} Corkum PB, 1993 {\it \PRL} {\bf71} 1994-7; Schafer KJ, Yang B, DiMauro LF and Kulander KC,
1993 {\it \PRL}{\bf70}, 1599-1602 


\bibitem{Volkovkick}Becker A and Faisal FHM 2000 {\it \PRL} {\bf84}, 3546; Chen J, Liu J, Fu LB and Zheng WM 2000 {\it Phys. Rev. A} {\bf63}, 011404


\bibitem{ChenandNam}Chen J and Nam CH 2002 {\it Phys. Rev. A} {\bf 66} 053415 

\bibitem{Z-NZ}Ho PJ and Eberly JH,
2003 {\it Optics Express} {\bf 11} 2826-2831;
Ho PJ, Panfili R, Haan SL, and Eberly JH 2005 {\it \PRL} {\bf94} 093002

\bibitem{RESI}Rudenko A {\it et al} 2004 {\it \PRL} {\bf 93} 253001


\bibitem{randomkick} Feuerstein B, Moshammer R, and Ullrich J
2000 {\it J. Phys. B: At. Mol. Opt. Phys.} {\bf33} L823-L830; Kopold R, Becker W, Rottke H, and Sandner W 2000 {\it \PRL} {\bf85} 3781
   

\bibitem{PRL2006} Haan SL, 
Breen L, Karim A, and Eberly JH 2006
\PRL{\bf97} 103008;
2007 {\it Optics Express}  {\bf15} 767 


\bibitem{ZachPRA}Haan SL and Smith ZS 
2007 {\it Phys. Rev. A} {\bf 76} 053412 


\bibitem{PRL2008}
Haan SL, Smith ZS, and VanDyke JS, 
2008 {\it \PRL}{\bf101} 113001 

\bibitem{hotelectrons}B. Witzel, N.A. Papadogiannis, and D. Charalambidis, 2000 {\it \PRL}{\bf85} 2268

\bibitem{parker390}Parker JS, Doherty BJS, Taylor KT, Schultz KD, Blaga CI and  DiMauro LF
2006 {\it \PRL} {\bf96} 133001

\bibitem{outsidebox-expt}Staudte A  {\it et al},
2007 {\it \PRL} {\bf 99} 263002;

Rudenko A {\it et al.},
2007{\it \PRL} {\bf 99} 263003

\bibitem{outsidebox-thy}
Emmanouilidou A 2008 {\it Phys. Rev. A} {\bf 78} 023411

Ye DF, Liu X, and Liu J 2008 {\it \PRL}{\bf101} 233003 



\bibitem{JPB2008}Haan SL, Smith ZS, Shomsky KN, and Plantinga PW
 2008 {\it J. Phys. B: At. Mol. Opt. Phys.} {\bf41} 211002 

\bibitem{wavelengths}Alnaser AS {\it et al} 2008 {\it J. Phys. B: At. Mol. Opt. Phys.} {\bf41} 031001 


\bibitem{panfili}
Panfili R, Haan SL, and Eberly JH, 
2002 {\it \PRL}{\bf89} 113001

\bibitem{RudenkoPRA2008}
Rudenko A {\it et al} 2008 {\it Phys.~Rev.~A} {\bf78} 015403










 


 

%

%

\end{thebibliography}
\end{document}